\documentclass[prb,superscriptaddress,twocolumn,longbibliography]{revtex4-2}

\usepackage{amsmath,amssymb}
\usepackage{physics}
\usepackage{graphicx}
\usepackage{bm}
\usepackage{bbm}
\usepackage[colorlinks,bookmarks=false,citecolor=blue,linkcolor=black,urlcolor=red]{hyperref}
\usepackage{color} \usepackage{srcltx} 
\usepackage{newtxtext,newtxmath}

\usepackage{ulem}

\allowdisplaybreaks

\begin{document}

\title{Chirality-selective proximity effect
between chiral $p$-wave superconductors and quantum Hall insulators
}
\date{\today}
\author{Ryota Nakai}
\affiliation{RIKEN Center for Quantum Computing (RQC), Wako, Saitama, 351-0198, Japan}

\author{Koji Kudo}
\affiliation{Department of Physics, Kyushu University, Fukuoka 819-0395, Japan}

\author{Hiroki Isobe}
\affiliation{Department of Physics, Kyushu University, Fukuoka 819-0395, Japan}

\author{Kentaro Nomura}
\affiliation{Department of Physics, Kyushu University, Fukuoka 819-0395, Japan}

\begin{abstract}
    Heterostructures of superconductors and quantum-Hall insulators are promising platforms of topological quantum computation.
    However, these two systems are incompatible in some aspects such as a strong magnetic field, the Meissner effect, and chirality.
    In this work, we address the condition that the superconducting proximity effect works in the bulk of quantum Hall states,
    and identify an essential role played by the vortex lattice regardless of pairing symmetry.
    We extend this finding to a heterostructure of a chiral $p$-wave superconductor in the mixed state and an integer quantum Hall insulator.
    The proximity effect works selectively in the lowest Landau level depending on relative chiralities.
    If the chiralities align, a topological phase transition to a topological superconductor occurs.
\end{abstract}

\maketitle

\section{Introduction}

Topology and pairing are the key to the creation of non-Abelian anyons.
Specifically, Majorana zero modes \cite{PhysRevLett.86.268,Alicea_2012,annurev:/content/journals/10.1146/annurev-conmatphys-030212-184337}, an example of the non-Abelian anyons, appear in e.g., 
superfluid $^3$He \cite{10.1093/acprof:oso/9780199564842.001.0001},
the Moore-Read fractional quantum Hall (QH) state \cite{MOORE1991362,PhysRevLett.66.802},
unconventional superconductors (SCs) \cite{PhysRevB.61.10267,AYuKitaev_2001}, and
quantum spin liquids \cite{KITAEV20062}.
All these systems have a non-trivial topology and intrinsic pairing of atoms, composite fermions \cite{PhysRevLett.63.199}, electrons, and emergent fermions, respectively.
On the other hand, it is also possible to extrinsically induce pairing by the superconducting proximity effect by making a heterostructure with SCs.
There are a variety of proposals for creating Majorana zero modes in heterostructures of $s$-wave SCs with e.g.,
topological insulators \cite{PhysRevLett.100.096407,PhysRevLett.102.216404,PhysRevB.79.161408},
quantum anomalous Hall (QAH) insulators \cite{PhysRevB.82.184516},
Rashba nanowires \cite{AYuKitaev_2001,PhysRevLett.104.040502,PhysRevLett.105.077001,PhysRevLett.105.177002,doi:10.1126/science.1222360}, and
magnetic atomic chains \cite{PhysRevB.84.195442,PhysRevB.88.020407,PhysRevLett.111.147202,PhysRevLett.111.186805,PhysRevLett.111.206802,PhysRevB.88.155420,PhysRevB.88.180503,doi:10.1126/science.1259327}.

The proximity effect creates a correlation between electrons and holes through penetrating Cooper pairs from SCs.
However, a proximitized SC does not necessarily induce the proximity effect.
This is because, in addition to microscopic details of the interface, the proximity effect is subject to the limitation as to whether electronic states in the non-SC side can accommodate Cooper pairs.
Specifically, singlet Cooper pairs cannot penetrate spin-polarized materials such as half metals due to the spin configuration \cite{PhysRevLett.86.4096,RevModPhys.77.1321,Eschrig2007}, 
nor can they penetrate magnetic Weyl semimetals due to the opposite chirality of the nodes, known as the chirality blockade \cite{PhysRevB.96.035437}.
Bulk QH states are another example as we will explain in the following.

SC/QH heterostructures have been proposed as a platform of more exotic non-Abelian anyons such as parafermions and Fibonacci anyons \cite{PhysRevX.2.041002,Clarke2013,PhysRevX.4.011036,PhysRevX.4.031009}.
There are roughly two configurations of the heterostructure depending on whether an SC is attached to the QH edge, or it covers the entire bulk.
Theoretical and experimental studies so far have paid particular attention to the former configuration with
integer \cite{ma93,zyuzin94,fisher94,PhysRevB.53.1548,ishikawa99,giazotto05,stone11,
lian16,
PhysRevB.94.064516,
PhysRevB.96.140506,
PhysRevB.96.241104,
PhysRevB.103.184509,
PhysRevB.104.115435,
10.21468/SciPostPhysCore.5.3.045,
PhysRevB.106.245411,
PhysRevB.107.125416,
PhysRevX.13.031027,
Kurilovich2023,
PhysRevResearch.5.013066,
PhysRevB.109.064519,
doi:10.1126/science.aad6203,
Lee2017,
Park2017,
doi:10.1126/sciadv.aaw8693,
Zhao2020,
doi:10.1021/acs.nanolett.2c01413,
Vignaud2023,
Barrier2024} 
and fractional \cite{PhysRevX.2.041002,Clarke2013,PhysRevX.4.011036,PhysRevX.4.031009,PhysRevX.12.021057,PhysRevLett.129.037703,PhysRevB.107.L161105} QH edges since the edge is the only conduction channel.
In contrast, for the latter configuration, 
there are possibly three factors that spoil the functionality of SC/QH heterostructures.
(i) QH states requires a strong magnetic field, which is likely to break superconductivity \cite{PhysRevB.99.094509,doi:10.1073/pnas.2202948119}.
(ii) Even if superconductivity is retained, the Meissner effect repels the magnetic field making QH states difficult to realize. 
(iii) If pairing is $s$-wave, Cooper pairs are formed between time-reversal pairs of electrons while QH states break time-reversal symmetry.
Nonetheless, it has been shown that a bulk QH state coupled with a mixed-state $s$-wave pair potential shows a topological phase transition to topological SCs \cite{PhysRevB.93.214504,PhysRevB.99.115427,PhysRevB.101.024516,PhysRevB.110.035147}.

The purpose of this study is twofold: to elucidate the condition that the $s$-wave and $p$-wave proximity effect works in the bulk QH states,
and to study the topological properties of a heterostructure of a mixed-state chiral $p$-wave SC and a QH insulator.
As for the former purpose,
we will show that the angular momentum of the Cooper pairs generated by the vortex lattice
is necessary to pair the bulk QH states.
This consequence is true regardless of whether pairing symmetry is $s$-wave or $p$-wave.
Though the $s$-wave case has been studied in Refs.~\onlinecite{PhysRevB.99.115427,PhysRevB.101.024516}, 
we will recast their study on the disk geometry to make the chiral nature of the heterostructure explicit and then extend the argument to the $p$-wave case.
Based on this finding, we consider a heterostructure of a mixed-state chiral $p$-wave SC and a QH insulator.
The two systems in this heterostructure have different kinds of chirality, that is, the sign of the Cooper pairs' angular momentum which is $\pm\hbar$ for $p_x\pm ip_y$ wave and the chirality of the Landau levels.
The proximity effect works selectively depending on the relative alignment of the chiralities
particularly in the lowest Landau level (LLL).
With an effective chiral $p$-wave pair potential, the LLL shows a topological phase transition to a topological SC. 

Notice that a chiral $p$-wave SC is itself a topological SC.
However, we can distinguish between the topological superconductivity of a chiral $p$-wave SC and that induced in a QH insulator (we will show this in Sec.~\ref{sec:chernnumber}).
For the purpose of creating more exotic anyons, the latter topological superconductivity is preferred since some proposals require the combination of fractional charges and pairing \cite{PhysRevX.2.041002,Clarke2013,PhysRevX.4.011036,PhysRevX.4.031009}.
The emergence of the proximity-induced topological superconductivity in an integer QH insulator is the first step toward this direction.

The rest of this paper is organized as follows.
In Sec.~\ref{sec:preliminaries}, we first give a phenomenological argument on an SC/QH heterostructure in a magnetic field, and show that, by using a continuum model on a disk, an impractical setup with $s$-wave SCs does not result in the bulk proximity effect.
Then, we recast the works in Refs.~\onlinecite{PhysRevB.99.115427,PhysRevB.101.024516} on a disk geometry and show the necessity of the vortex lattice for the $s$-wave case. 
In Sec.~\ref{sec:continuous}, we show that the same problem arises with chiral $p$-wave SCs, but is resolved by mixed-state SCs.
In Sec.~\ref{sec:lattice}, we confirm these results using a tight-binding model.
Finally, we conclude in Sec.~\ref{sec:conclusion}.

\section{Preliminaries}
\label{sec:preliminaries}

\subsection{SC/QH insulator heterostructure}
\label{sec:generalhetero}

Throughout this paper, we consider heterostructures where the bulk of a two-dimensional electron gas is covered by an SC and a strong perpendicular magnetic field is applied.

\begin{figure}
    \centering
    \includegraphics[width=0.47\textwidth]{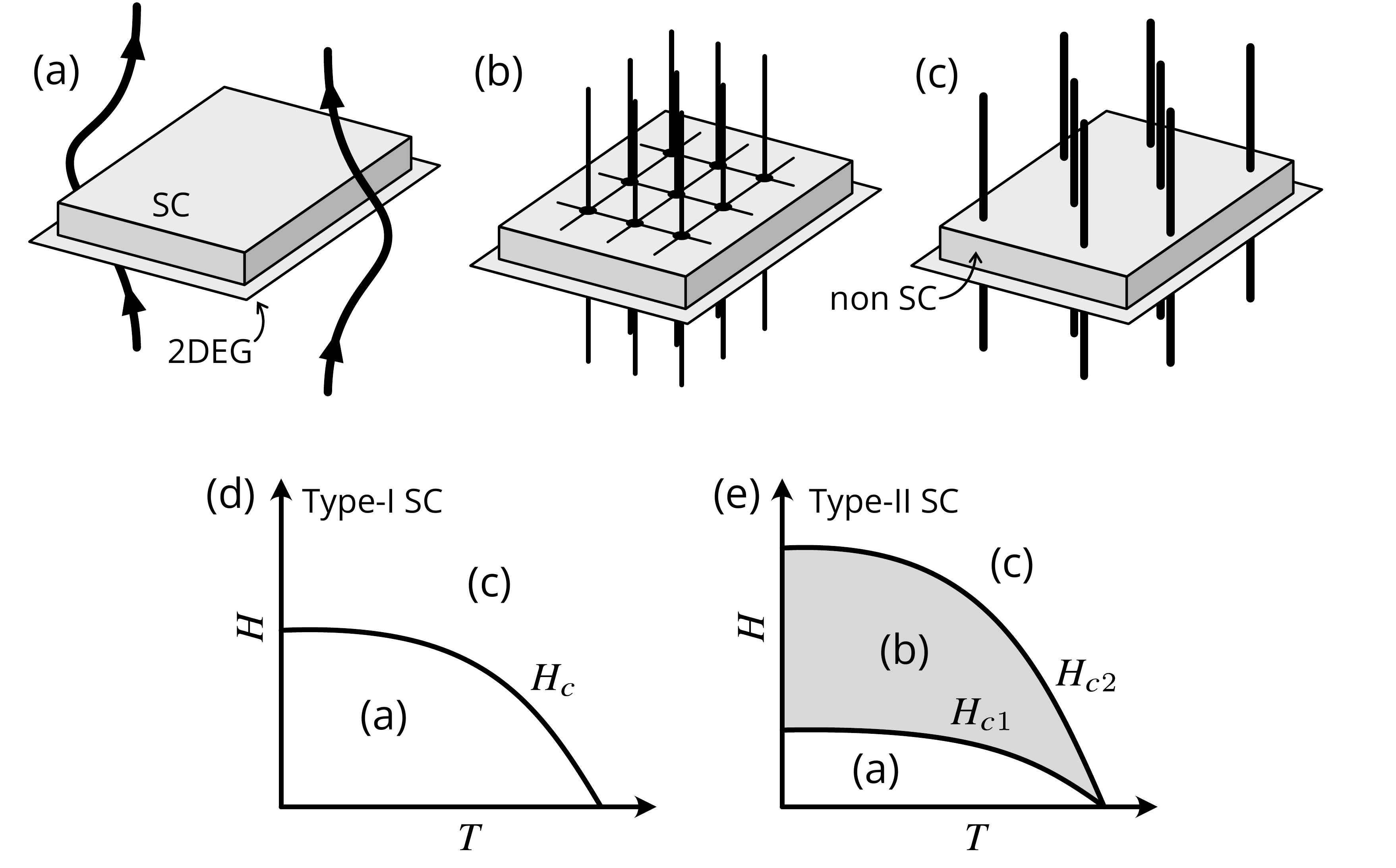}
    \caption{
        A heterostructure of a type-I or type-II superconductor and a quantum Hall insulator
        (a) below $H_c$ or $H_{c1}$ where the magnetic field is completely screened by the Meissner effect,
        (b) in the mixed state between $H_{c1}$ and $H_{c2}$ where the vortex lattice is formed, and
        (c) above $H_c$ or $H_{c2}$ where the superconductivity is broken.
        (d), (e) The corresponding phases in the phase diagrams of type-I and type-II superconductors, respectively.
        \label{fig:schematic}
    }
\end{figure}

First, we focus on the SC side \cite{tinkham2004introduction}.
An SC subject to a strong magnetic field is in one of the following three cases:
(a) below $H_c$ in type-I SCs or below $H_{c1}$ in type-II SCs, the magnetic field is screened by the Meissner effect, 
(b) between $H_{c1}$ and $H_{c2}$ in type-II SCs, the magnetic field penetrates the SC as a vortex lattice, or
(c) above $H_c$ in type-I SCs or above $H_{c2}$ in type-II SCs, the superconductivity is lost (Fig.~\ref{fig:schematic}).
In each phase, the pair potential $\Delta(\bm{r})$ and the magnetic field $B(\bm{r})$ is 
(a) $|\Delta(\bm{r})|=\Delta_0$ and $B(\bm{r})=0$,
(b) $|\Delta(\bm{r})|< \Delta_0$ but nonzero and $B(\bm{r})\neq 0$,
(c) $|\Delta(\bm{r})|=0$ and $B(\bm{r})= H$,
where $\Delta_0$ is the pair potential of an SC below $H_c$ or $H_{c1}$.
We here set the permeability $\mu_0=1$.

As for the two-dimensional electron gas side, we need both the strong magnetic field and the proximity effect to construct an SC/QH insulator heterostructure.
Either one of them is lost in the cases (a) and (c).
The remaining possibility is the case (b).
However, even in the presence of both the magnetic field and the superconductivity, it is nontrivial whether the proximity effect works in QH insulators.
One of the main claim of this work is that the case (b) is essential regarding not just the presence of both superconductivity and a magnetic field but regarding the proximity effect as well.
In Sec.~\ref{sec:mixeds_disk} and \ref{sec:continuous}, we will discuss the condition of the presence of the proximity effect in QH states from the viewpoint of the angular momentum.
Before that, we will review in the next subsection that a uniform pair potential in a QH insulator, which would be the simplest theoretical assumption but is not listed in Fig.~\ref{fig:schematic}, fails to induce the bulk proximity effect.

\subsection{Uniform $s$-wave pair potential}
\label{sec:uniforms_disk}

The Hamiltonian of a continuum model of a spinful but spin-unpolarized QH insulator on a disk that couples with an $s$-wave pair potential $\Delta(\bm{r})$ is given by
$
    H^s = H^\text{spinful}_0 + H^s_\Delta
$,
where
\begin{align}
    H^\text{spinful}_0
    &=
    \sum_{\sigma=\uparrow,\downarrow}
    \int d\bm{r}
    \psi^\dagger_\sigma(\bm{r})
    h_0(\bm{r})
    \psi_\sigma(\bm{r}),
    \label{eq:normalstatehamiltonian_spinful}\\
    H^s_\Delta
    &=
    \int d\bm{r}
    \psi^\dagger_\uparrow(\bm{r})
    \Delta(\bm{r})
    \psi^\dagger_\downarrow(\bm{r})
    +
    \text{H.c.}
    \label{eq:pairinghamiltonian_s}
\end{align}
and 
\begin{align}
    h_0(\bm{r})
    =
    \frac{(-i\hbar\bm{\partial}+e\bm{A})^2}{2m_e}-\mu,
\end{align}
where $m_e$ is the electron mass, $-e<0$ is the electron charge, $\mu$ is the chemical potential, and we employ the symmetric gauge $\bm{A}=(-By/2,Bx/2)$.
$\psi_\sigma(\bm{r})$ is the electron annihilation operator of spin $\sigma$. 
We assume that the Zeeman term is negligible due to a small $g$-factor.
The simultaneous eigenstate of the normal-state Hamiltonian $h_0(\bm{r})$ with eigenvalue $\hbar\omega_c(n+1/2)\,(n\in\mathbb{Z}, n\ge 0,\omega_c=eB/m_e)$ and the angular momentum operator with eigenvalue $-\hbar m\,(m\in\mathbb{Z}, m\ge -n)$ is given by the Landau level wavefunction \cite{Jain07}
\begin{align}
    &\phi_{nm}(\bm{r})
    =
    \frac{{a^{\dagger}}^n{b^\dagger}^{n+m}}{\sqrt{n!(n+m)!}}\phi_{00}(\bm{r}) \notag\\
    &=
    \frac{(-1)^n}{\sqrt{2\pi}\ell_B}
    \sqrt{\frac{n!}{2^m(n+m)!}}\left(\frac{z}{\ell_B}\right)^me^{-z\bar{z}/4\ell_B^2}L_n^m\left(\frac{z\bar{z}}{2\ell_B^2}\right),
    \label{eq:landaulevel_disk}
\end{align}
where $\ell_B=\sqrt{\hbar/eB}$, $\phi_{00}=e^{-z\bar{z}/4\ell_B^2}/\sqrt{2\pi}\ell_B$, $z=x-iy$, $L_n^m$ is the associated Laguerre polynomial, and
\begin{align}
    &a=\frac{1}{\sqrt{2}}\left(2\ell_B\partial_{\bar{z}}+\frac{z}{2\ell_B}\right),\,
    &b=\frac{1}{\sqrt{2}}\left(2\ell_B\partial_{z}+\frac{\bar{z}}{2\ell_B}\right).
\end{align}
On the Landau level basis $\phi_{nm}(\bm{r})$ with $\psi_\sigma(\bm{r}) = \sum_{nm} \phi_{nm}(\bm{r}) c_{nm\sigma}$, the pairing Hamiltonian is written as
\begin{align}
    H_\Delta^s
    &=
    \sum_{nmn'm'}\Delta_{nmn'm'}^sc_{nm\uparrow}^\dagger c_{n'm'\downarrow}^\dagger
    +
    \text{H.c.},
\end{align}
where the matrix element is
\begin{align}
    \Delta_{nmn'm'}^s
    =
    \int d\bm{r}\phi_{nm}^\ast(\bm{r})\Delta(\bm{r})\phi_{n'm'}^\ast(\bm{r}).
    \label{eq:pairpotential_matrixelement_sdisk}
\end{align}
Since $a^{\ast}=b$, the complex conjugation of $\phi_{nm}$ can be obtained by interchanging $a^\dagger$ and $b^\dagger$ in (\ref{eq:landaulevel_disk}), which gives $\phi_{nm}^\ast=\phi_{n+m,-m}$.
With this identity, the matrix element for a uniform pair amplitude $\Delta(\bm{r})=\Delta_0$ gives
\begin{align}
    \Delta_{nmn'm'}^s
    =
    \Delta_0\delta_{n,n'+m'}\delta_{n+m,n'}.
    \label{eq:pairpotential_matrixelement_sdisk_result}
\end{align}

Projecting onto the LLL by $n=n'=0$, the pairing Hamiltonian reduces to
\begin{align}
    H^s_\Delta
    \to
    \Delta_0c_{00\uparrow}^\dagger c_{00\downarrow}^\dagger
    +
    \text{H.c.}.
\end{align}
This implies that only spin-up and -down electrons with angular momentum $0$ can make a Cooper pair by the proximity effect from a uniform $s$-wave SC.
The same conclusion holds also for higher Landau levels if pairing is restricted within a single Landau level.

This result can be understood from the perspective of the angular momentum.
The angular momentum of an $s$-wave Cooper pair is $0$ and that of the LLL state electrons is 0 or negative.
As a result, possible pairing is between electrons with angular momentum 0.
Indeed, since $\phi_{0m}^\ast\phi_{0m'}^\ast$ contains $\bar{z}^{m+m'}$,
the angular integral in (\ref{eq:pairpotential_matrixelement_sdisk}) gives 0 unless $m=m'=0$.
Macroscopically, the amount of induced Cooper pairs in the LLL is negligible in the thermodynamic limit, and hence
we cannot anticipate topological phase transitions.

In addition, the above proximity effect breaks spatial translation symmetry.
Specifically, the LLL state with $m$ is distributed around a circle of radius $\sqrt{2m}\ell_B$.
This indicates that the uniform $s$-wave pair potential works only at the center ($m=0$) of the disk, which is compatible with the vanishing proximity effect in the thermodynamic limit.
Similar things happen in different gauges and geometries.
With the Landau gauge and the cylinder geometry, the uniform pair potential works dominantly for $k=0$ states (Appendix \ref{app:cylinder}),
and with the Dirac monopole's gauge and the spherical geometry, it works at the north and south poles.
In our setup, this problem is considered to be resulting from an impractical setup, that is, a uniform $s$-wave pair potential is applied in a QH insulator, which is not listed in Fig.~\ref{fig:schematic}.

\subsection{Mixed-state $s$-wave pair potentials}
\label{sec:mixeds_disk}

In a mixed-state SC, quantized magnetic fluxes $h/2e$ form the Abrikosov lattice. 
Each flux is screened by supercurrent flowing around a vortex, which results in a center-of-mass angular momentum of the Cooper pairs.
The Cooper pairs can have a variety of total angular momentum, which is the summation of the relative and center-of-mass ones. This removes the constraint by the angular momentum in QH insulators.
The $s$-wave case corresponds to Refs.~\onlinecite{PhysRevB.93.214504,PhysRevB.99.115427,PhysRevB.101.024516,PhysRevB.110.035147}.
We review their system from the viewpoint of the angular momentum.
Notice that, in the following, we assume that the magnetic field is spatially uniform due to a long penetration depth compared with the distance of neighboring vortices slightly below $H_{c2}$. 

The pair potential $\Psi(\bm{r})$ of SCs close to $H_{c2}$ obeys the linearized Ginzburg-Landau equation \cite{tinkham2004introduction}
\begin{align}
    \left[\frac{(-i\hbar\bm{\partial}+e^\ast \bm{A})}{2m^\ast_e}+\alpha\right] \Psi(\bm{r}) = 0,
\end{align}
where $e^\ast=2e$, $m^\ast_e=2m_e$, and $\alpha<0$.
This equation is the same as that of electrons except that the mass and charges are those of Cooper pairs.
The pair potential is written in terms of the LLL wavefunctions of Cooper pairs.
Assuming that the pair potential $\Delta(\bm{r})$ induced in a QH insulator is proportional to $\Psi(\bm{r})$, we have
\begin{align}
    \Delta(\bm{r})
    =
    \sum_{m\ge 0}\frac{C_m}{\sqrt{\pi m!}\ell_B}
    \left(\frac{z}{\ell_B}\right)^me^{-z\bar{z}/2\ell_B^2}.
    \label{eq:mixedstatepairpotential_disk}
\end{align}
Each wavefunction is characterized by the angular momentum $-m\hbar$, which is distributed around a circle of radius $\sqrt{m}\ell_B$.
The magnitude of $C_m$ is approximately $|C_m|\simeq \ell_B\sqrt{\pi\langle|\Delta(\bm{r})|^2\rangle}$ irrespective of $m$, where $\langle\cdots\rangle$ denotes the spatial average (Appendix \ref{sec:pairpotential_disk}). 
We note that $\Delta(\bm{r})$ is slowly varying in the vortex lattice scale slightly below $H_{c2}$.
The pair potential (\ref{eq:mixedstatepairpotential_disk}) contains $m_0$ vortices when the summation of the right-hand side is truncated to $m_0$.
However, the position of the vortices depends on $C_m$, which is determined to minimize the Ginzburg-Landau free energy \cite{tinkham2004introduction}.

QH states can form an extensive number of Cooper pairs by (\ref{eq:mixedstatepairpotential_disk}).
Specifically, the matrix element between the LLL states is given by (Appendix \ref{sec:pairpotential_disk})
\begin{align}
    \Delta_{0m0m'}^s
    =
    \frac{C_{m+m'}}{2\sqrt{\pi}\ell_B}
    \sqrt{\frac{(m+m')!}{2^{m+m'}m!m'!}}.
    \label{eq:matrixelement_mixedstatespairpotential_disk}
\end{align}
The matrix element (\ref{eq:matrixelement_mixedstatespairpotential_disk}) with a fixed $m+m'$ is maximized when $|m-m'|$ is minimized.
This is because there is a substantial overlap between the Cooper pair wavefunction with angular momentum $-2m\hbar$ and the electronic LLL wavefunction with angular momentum $-m\hbar$
since they are distributed along the same circle of radius $\sqrt{2m}\ell_B$.
In addition, the angular dependence of $\bar{z}^{2m}$ in $\phi_{0m}^\ast\phi_{0m}^\ast$ is cancelled by that of $z^{2m}$ in $\Delta(\bm{r})$, which makes the integral (\ref{eq:pairpotential_matrixelement_sdisk}) nonvanishing.
The mixed-state pair potential works as a glue to the Landau level states.
A similar discussion at a QH edge coupled with a Rashba SC has been given in Ref.~\cite{PhysRevResearch.5.013066}.

\section{Continuum model}
\label{sec:continuous}

In this section, we identify the necessity of the vortex lattice by considering the proximity effect from
a uniform and mixed-state chiral $p$-wave pair potentials, where time-reversal symmetry is broken as in QH insulators.

\subsection{A relation between $s$-wave and chiral $p$-wave pair potentials}

First, we derive a general relation between the matrix elements of the $s$-wave and chiral $p$-wave pair potentials.
The Hamiltonian of a spinless QH insulator coupled with a spinless chiral $p$-wave pair potential is given by
$
    H^{p_x\pm ip_y} = H^\text{spinless}_0 + H^{p_x\pm ip_y}_\Delta
$,
where
\begin{align}
    H^\text{spinless}_0
    &=
    \int d\bm{r}
    \psi^\dagger(\bm{r})
    h_0(\bm{r})
    \psi(\bm{r}),
    \label{eq:normalstatehamiltonian_spinless}\\
    H^{p_x\pm ip_y}_\Delta
    &=
    \frac{1}{2}
    \int d\bm{r}
    \psi^\dagger(\bm{r})
    \frac{\left\{-i\hbar\partial_\pm,\Delta(\bm{r})\right\}}{2}
    \psi^\dagger(\bm{r})
    +
    \text{H.c.},
    \label{eq:pairinghamiltonian_chihralp}
\end{align}
and $\partial_\pm\equiv\partial_x\pm i\partial_y$.
$\psi(\bm{r})$ is the electron annihilation operator.
On the Landau level basis, the pairing Hamiltonian is rewritten as
\begin{align}
    H_\Delta^{p_x\pm ip_y}
    &=
    \sum_{nmn'm'}\Delta_{nmn'm'}^{p_x\pm ip_y}c_{nm}^\dagger c_{n'm'}^\dagger
    +
    \text{H.c.}.
\end{align}
We can replace derivatives in (\ref{eq:pairinghamiltonian_chihralp}) by the covariant derivatives as
\begin{align}
    \Delta_{nmn'm'}^{p_x\pm ip_y}
    =
    \frac{1}{2}
    \int d\bm{r}
    \phi_{nm}^\ast
    \frac{-i\hbar D_{\pm}^h\Delta+\Delta\left(-i\hbar D_{\pm}^e\right)}{2}
    \phi_{n'm'}^\ast,
    \label{eq:pairpotential_matrixelement_chiralp}
\end{align}
where $-i\hbar D_\alpha^{e/h}=-i\hbar\partial_\alpha\pm eA_\alpha$ ($\alpha=\pm$, $A_\pm\equiv A_x\pm iA_y$) is the covariant derivative for electrons and holes, respectively.
Different types of the covariant derivatives appear in (\ref{eq:pairpotential_matrixelement_chiralp}) due to the gauge invariance.
Each term of (\ref{eq:pairpotential_matrixelement_chiralp}) is invariant under a gauge transformation
\begin{align}
    \phi_{nm}(\bm{r})&\to e^{ie\chi(\bm{r})/\hbar}\phi_{nm}(\bm{r}), \notag\\
    \Delta(\bm{r})&\to e^{2ie\chi(\bm{r})/\hbar}\Delta(\bm{r}), \notag\\
    \bm{A}(\bm{r})&\to \bm{A}(\bm{r})-\bm{\partial}\chi(\bm{r}). \notag
\end{align}
Since $\phi_{nm}^\ast$ and $\phi_{n'm'}^\ast$ transform like holes, while $\phi_{nm}^\ast\Delta$ and $\Delta\phi_{n'm'}^\ast$ transform like electrons, the derivative of the first (second) term should transform like that for holes (electrons).
The electron covariant derivative acts like raising or lowering operator as
\begin{align}
    -i\hbar D_{\pm}^e\phi_{nm}
    =
    \frac{\sqrt{2}i\hbar}{\ell_B}
    \left\{
        \begin{array}{ll}
            \sqrt{n+1}\phi_{n+1m-1} \, &(+)\\
            -\sqrt{n}\phi_{n-1m+1} \, &(-)\\
        \end{array}
    \right.,
\end{align} 
where we define $\phi_{nm}=0$ with $n<0$.
With this result, the matrix elements of the $s$-wave and chiral $p$-wave pair potentials with a common $\Delta(\bm{r})$ are related by
\begin{align}
    &\Delta_{nmn'm'}^{p_x\pm ip_y}
    =
    -\frac{i\hbar}{2\sqrt{2}\ell_B} \times \notag\\
    &
    \left\{
        \begin{array}{ll}
            -\sqrt{n}\Delta_{n-1m+1n'm'}^s+\sqrt{n'}\Delta_{nmn'-1m'+1}^s \, &(p_x+ip_y)\\
            \sqrt{n+1}\Delta_{n+1m-1n'm'}^s-\sqrt{n'+1}\Delta_{nmn'+1m'-1}^s \, &(p_x-ip_y)\\
        \end{array}
    \right..
    \label{eq:pairpotential_schiralprelation_disk}
\end{align} 
Here, $\Delta_{nmn'm'}^s$ is given in (\ref{eq:pairpotential_matrixelement_sdisk}).

\subsection{Uniform chiral $p$-wave pair potential}

The absence of the bulk proximity effect from uniform SCs is true even if we consider a chiral $p$-wave pair potential.
For a uniform chiral $p$-wave pair potential $\Delta(\bm{r})=\Delta_0$, one can evaluate the matrix element by (\ref{eq:pairpotential_matrixelement_sdisk_result}) and (\ref{eq:pairpotential_schiralprelation_disk}).
Projecting onto the LLL, we have
\begin{align}
    H^{p_x\pm ip_y}_\Delta
    \to
    \left\{
        \begin{array}{ll}
            0 \, &(p_x+ip_y) \\
            \displaystyle \frac{i\hbar\Delta_0}{\sqrt{2}\ell_B} c_{01}^\dagger c_{00}^\dagger + \text{H.c.} \, &(p_x-ip_y)
        \end{array}
    \right..
\end{align}
This implies that the proximity effect does not work in the LLL from $p_x+ip_y$-wave SCs, 
while that from $p_x-ip_y$ SCs couples electrons with angular momentum $m=0$ and $1$.
Since the relative angular momentum of the Cooper pairs is $\pm\hbar$ in $p_x\pm ip_y$-wave SCs,
the pairing between $m=0$ and $1$ is possible by the $p_x-ip_y$-wave pair potential.
Importantly, the proximity effect from uniform SCs regardless of whether pairing symmetry is $s$-wave or $p$-wave does not induce bulk proximity effect in QH insulators, and is negligible in the thermodynamic limit.

\subsection{Mixed-state chiral $p$-wave pair potentials}

What is new by considering a mixed-state chiral $p$-wave pair potential is that the pair potential on the LLL depends on the relative chirality.
The matrix element of a mixed-state chiral $p$-wave pair potential in the LLL is readily obtained by (\ref{eq:pairpotential_schiralprelation_disk}) given the function $\Delta(\bm{r})$ is the same as the $s$-wave case [Eq.~\eqref{eq:mixedstatepairpotential_disk}].
As in the uniform case, the $p_x+ip_y$-wave pair potential does not work in the LLL ($\Delta_{0m0m'}^{p_x+ ip_y}=0$), while the $p_x-ip_y$-wave pair potential is given by (see Appendix \ref{sec:pairpotential_disk} for derivation)
\begin{align}
    \Delta_{0m0m'}^{p_x- ip_y}
    &=
    C_{m+m'-1}\frac{i\hbar (m-m')}{4\sqrt{\pi}\ell_B^2}
    \sqrt{\frac{(m+m'-1)!}{2^{m+m'}m!m'!}}.
    \label{eq:matrixelement_mixedstatechiralppairpotential_disk}
\end{align} 
This is one of the main result of this work.
While LLL states with the same $m$ cannot make pairs due to the Pauli's exclusion principle (we assumed spinless chiral $p$-wave pairing), 
those with smaller $|m-m'|$ but $m \neq m'$ are likely to form Cooper pairs.

From the perspective of the angular momentum, the chirality dependence is not obvious.
As the center-of-mass angular momentum is the same for all pairing symmetries,
the presence and absence of the proximity effect on the LLL is not restricted by the angular momentum but the chirality of the $p$-wave pair potential. 
Indeed, the proximity effect from both chirality do work in higher Landau levels.
We will examine the above result in a lattice model in Sec.~\ref{sec:lattice}.
Notice that it is one of the advantage of considering the disk geometry that the chirality of the pairing and that of the center-of-mass motion can be discussed on the same footing.

In our heterostructure, distinct chiralities are attributed to three types of motions, that is, the center-of-mass motion of Cooper pairs determined by the Ginzburg-Landau equation, the relative motion of Cooper pairs which is $\pm\hbar$ for $p_x\pm ip_y$ wave, and that of QH states. 
Suppose we flip the direction of the applied magnetic field, 
the chirality of the center-of-mass motion and the QH states are flipped by definition.
On the other hand, the chirality of pairing symmetry are not strictly tied to the applied magnetic field, but it depends energetically on the direction of the field \cite{PhysRevB.92.134520}.
As a result, the same phenomenon with the opposite chirality would occur by flipping the direction of the magnetic field.

\subsection{Chern number}
\label{sec:chernnumber}

Before we proceed to the lattice model, we review our heterostructures from the perspective of the Chern number.
The purpose of this subsection is to show that the emergence of the topological superconductivity in a chiral $p$-wave SC/QH insulator heterostructure is a consequence of a topological phase transition in the QH insulator.
We can distinguish between the topological superconductivity intrinsic to a chiral $p$-wave SC and that induced in a QH insulator.
Throughout this paper, we use the Bogoliubov-de Gennes (BdG) Chern number
$\mathcal{N}$ of the quasi-hole bands.
Here, we assume the periodic boundary condition in both the $x$ and $y$ directions.
Let the eigenspinor $\eta_{j\bm{k}}$ of the momentum-space BdG Hamiltonian $h(\bm{k})$ be specified by the band number $j$ and the momentum $\bm{k}$.
The BdG Chern number of a band $j$ is the integral of the Berry curvature over the Brillouin zone
\begin{align}
    \mathcal{N}_j
    =
    \frac{1}{2\pi}\int_\text{BZ}d\bm{k}\left[\nabla_{\bm{k}}\times \bm{A}_j(\bm{k})\right]_z,
\end{align}
where $\bm{A}_j(\bm{k})=-i\eta_{j\bm{k}}^\dagger\bm{\nabla}_{\bm{k}}\eta_{j\bm{k}}$.
The BdG Chern number $\mathcal{N}$ of the quasi-hole bands are defined by the summation of $\mathcal{N}_j$ over the negative-energy bands.
In non-superconducting systems, the BdG Chern number is twice the usual Chern number, e.g., $\mathcal{N}=2$ when the spin-polarized LLL is filled.

First, we review the BdG Chern number of the mixed-state $s$-wave SC/QH insulator heterostructure \cite{PhysRevB.99.115427,PhysRevB.101.024516,PhysRevB.110.035147}.
A mixed-state $s$-wave superconductor has $\mathcal{N}=0$ while a QH insulator has $\mathcal{N}=2$, and hence the total BdG Chern number without the proximity effect is $\mathcal{N}=2$.
With the proximity effect, the QH state can become a topological superconductor with odd $\mathcal{N}$.
The total BdG Chern number after the topological phase transition is, of course, odd.
This indicates the appearance of topological superconductivity (odd total $\mathcal{N}$) is attributed to the proximity-induced topological phase transition in a QH insulator.

A similar story holds in a mixed-state chiral $p$-wave SC/QH insulator heterostructure.
The caveat here is that mixed-state chiral $p$-wave SCs have even $\mathcal{N}$ \cite{PhysRevB.92.134519,PhysRevB.92.134520}.
Each isolated vortex in chiral $p$-wave SCs binds topologically a Majorana zero mode \cite{PhysRevB.61.10267}.
When two or more vortices are getting closer, the bound Majorana fermions start tunneling between them.
When vortices form a uniform lattice, the Majorana fermions form energy bands inside the superconducting energy gap.
What Refs.~\onlinecite{PhysRevB.92.134519,PhysRevB.92.134520} have showed was that the tight-binding model of tunneling Majorana fermions is written by a $\pi$-flux model,
and thus the gapped subband have odd $\mathcal{N}$.
Together with the chiral $p$-wave condensate's odd BdG Chern number, the mixed-state chiral $p$-wave SCs have even $\mathcal{N}$.
Therefore, the Chern-number argument for the mixed-state $s$-wave case also applies to the mixed-state chiral $p$-wave case,
that is, an odd total $\mathcal{N}$ is attributed to the proximity-induced topological phase transition in a QH insulator.
In other words, the topological SC phase in the chiral $p$-wave SC cannot be continuously connected to the induced one in a QH insulator.

Notice, however, that the Chern number of the mixed-state chiral $p$-wave SCs in Refs.~\onlinecite{PhysRevB.92.134519,PhysRevB.92.134520} has been evaluated away from $H_{c2}$.
Specifically, they used a pair potential $\Delta_0e^{i\theta(\bm{r})}$ that has spatially uniform absolute value and the phase $\theta(\bm{r})$ determined by the London equations
\begin{align}
    \bm{\partial}\times\bm{\partial}\theta(\bm{r})=2\pi\hat{z}\sum_j\delta(\bm{r}-\bm{r}_j),\quad
    \bm{\partial}^2\theta(\bm{r})=0,
\end{align} 
where $\bm{r}_j$ is the center of a vortex.
This form of the pair potential is valid when $H\ll H_{c2}$ where the distance between neighboring vortices is much longer than the coherence length. 
Since our focus is a region close to $H_{c2}$, we examine whether the conclusion in Refs.~\onlinecite{PhysRevB.92.134519,PhysRevB.92.134520} is true even near $H_{c2}$ in the next section.

\section{Lattice model}
\label{sec:lattice}

In this section, we examine the following three points using tight-binding models of two systems, a chiral $p$-wave SC and a chiral $p$-wave SC/QH insulator heterostructure:
(i) the BdG Chern number of a mixed-state chiral $p$-wave SC is even,
(ii) the proximity effect in the LLL depends on the chirality of the $p$-wave pair potential, and
(iii) a topological superconductivity appear in the heterostructure.
Notice that the following calculations are done under the periodic boundary condition in both $x$ and $y$ direction.
We confirmed the same conclusions for the continuum model on a cylinder as that on a disk (see Appendix \ref{app:cylinder}).

\subsection{Tight-binding model}

We consider a common square-lattice tight-binding Hamiltonian for the chiral $p$-wave SC and the QH insulator, which is given by
\begin{align}
    H 
    =&
    -\sum_{\bm{r}\bm{\delta}}t_{\bm{r}\bm{r}+\bm{\delta}}c_{\bm{r}}^\dagger c_{\bm{r}+\bm{\delta}}^{\ }
    -\mu\sum_{\bm{r}} c_{\bm{r}}^\dagger c_{\bm{r}}^{\ } \notag\\
    &+
    \frac{1}{2}\sum_{\bm{r}\bm{\delta}}\left(\Delta_{\bm{r}\bm{r}+\bm{\delta}}c_{\bm{r}}^\dagger c_{\bm{r}+\bm{\delta}}^\dagger + \text{H.c.}\right),
    \label{eq:tightbinding}
\end{align}
where $\bm{r}\in\mathbb{Z}^2$, $\bm{\delta}=\pm\hat{x}$ or $\pm\hat{y}$ and the lattice constant is set to 1.
The chiral $p$-wave pair potential is $\Delta_{\bm{r}\bm{r}+\bm{\delta}}=\Delta(\bm{r}+\bm{\delta}/2)s_{\bm{\delta}}$, where $s_{\pm\hat{x}}=\mp i$, $s_{\pm\hat{y}}=\pm C$, and the chirality $C=\pm 1$ corresponds to the $p_x\pm ip_y$ pairing.
The magnetic unit cell size is $N_x\times N_y$.
The nearest-neighbor hopping amplitude is subject to a magnetic flux $\phi_0/N_xN_y$ per plaquette, where $\phi_0=h/e$, and hence $t_{\bm{r}\bm{r}\pm\hat{x}}=t$ and $t_{\bm{r}\bm{r}\pm\hat{y}}=te^{\pm 2\pi ix/N_xN_y}$ under the Landau gauge used in \cite{PhysRevB.92.134519,PhysRevB.92.134520}.

\begin{figure}
    \centering
    \includegraphics[width=0.47\textwidth]{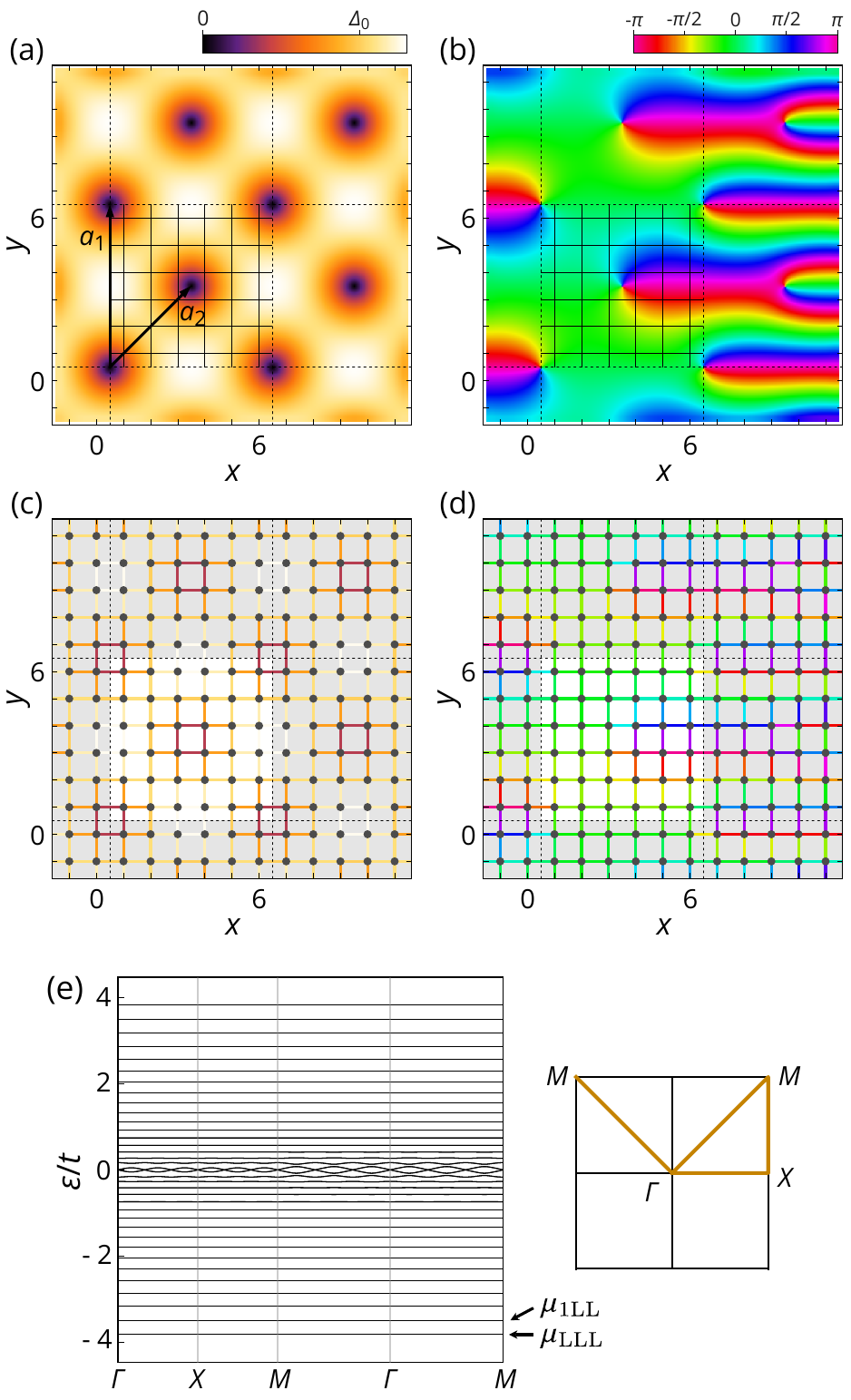}
    \caption{
        (a) The absolute value and (b) the phase of the pair potential in the continous space.
        Dashed lines represent the boundaries of a magnetic unit cell, and solid lines in a cell represent the lattice of the tight-binding model ($N_x=N_y=6$).
        (c) and (d) represent the absolute value and the phase of the pair potential, respectively, assigned to the nearest-neighbor bonds in the tight-binding model.
        The outside of a magnetic unit cell is shaded. Colors of bonds are represented in a common scale with the corresponding continuous ones.
                (e) Electronic band structure without pair potentials.
        The horizontal axis is along a line shown in the Brillouin zone.
        The lowest and the next lowest energy bands correspond to the lowest and first Landau levels (LLL and 1LL) whose energy is denoted by $\mu_\text{LLL}$ and $\mu_\text{1LL}$, respectively.
        \label{fig:pairpotential}
    }
\end{figure}

Notice that the meaning of the Hamiltonian (\ref{eq:tightbinding}) is different depending on whether we regard it as a model of a chiral $p$-wave SC or that of a chiral $p$-wave SC/QH insulator heterostructure.
For the former case, $c$ is the electron annihilation operator of the SC and $\Delta_{\bm{r}\bm{r}+\bm{\delta}}$ is the mean field of the BCS theory assuming $p$-wave pairing.
For the latter case, $c$ is that of the QH insulator and $\Delta_{\bm{r}\bm{r}+\bm{\delta}}$ is the pair potential induced in the QH insulator by the proximity effect from the chiral $p$-wave SC.
Notice also that even for the chiral $p$-wave SC, one needs to incorporate the magnetic field via the Peierls phase \cite{PhysRevB.92.134519,PhysRevB.92.134520}.
As a result, it is reasonable to consider the same Hamiltonian for both systems,
while the difference is the energy scale of $\mu$ and $\Delta_{\bm{r}\bm{r}+\bm{\delta}}$ compared with the hopping amplitude $t$, as will be specified later.
Without the pair potential, the model (\ref{eq:tightbinding}) shows a quantum Hall state where first several lowest energy bands is almost flat and have BdG Chern number 2 [Fig.~\ref{fig:pairpotential} (e)].
So, we call them as the Landau levels.

We consider a pair potential forming a square vortex lattice whose lattice vectors are $\bm{a}_1=(0,N_y)$ and $\bm{a}_2=(N_x/2,N_y/2)$ [Fig.~\ref{fig:pairpotential} (a) and (b)].
The continuous pair potential on a cylinder is rescaled to fit the tight-binding model and is shifted so that the vortex position is $(1/2,1/2)$ and $((N_x+1)/2,(N_y+1)/2)$ to avoid singularity on the lattice cites and bonds.
The resulting pair potential is given by (see Appendix \ref{app:pplatice})
\begin{align}
    \Delta(\bm{r})
    =
    \Delta_0
    &\sum_j
    e^{\pi ij^2/2}\exp\left[2\pi ij \frac{y-O_y}{N_y}\right] \notag\\
    &\times
    \exp\left[-2\pi\frac{(x-O_x+j N_x/2)^2}{N_xN_y}\right],
    \label{eq:pairpotential_tightbinding}
\end{align}
where $\bm{O}=(N_x/4+1/2,3N_y/4+1/2)$ [Fig.~\ref{fig:pairpotential} (c) and (d)].

The Hamiltonian at this moment is not invariant under translation by $\bm{a}_2$.
The translation invariance is recovered by introducing a singular gauge transformation \cite{PhysRevB.63.134509}
\begin{align}
    c_{\bm{r}}\to e^{i\text{arg}\Delta(\bm{r})/2}c_{\bm{r}}.
    \label{eq:singulargaugetransformation}
\end{align}
Since $\Delta(\bm{r}+\bm{a}_2)=e^{i\pi/2-2\pi iy/N_y}\Delta(\bm{r})$ 
and $t_{\bm{r}+2\bm{a}_2,\bm{r}+2\bm{a}_2\pm\hat{y}}=t_{\bm{r},\bm{r}\pm\hat{y}}e^{\pm 2\pi i/N_y}$
the translation invariance of the hopping amplitude is recovered.
Notice that the singular gauge transformation (\ref{eq:singulargaugetransformation}) have a branch cut,
so we introduce a branch cut line connecting neighboring two vortices \cite{PhysRevB.92.134520} so that branch cut lines do not cross the boundaries of the magnetic unit cells.
In the following we consider $N_x=N_y=6$ as shown in Fig.~\ref{fig:pairpotential} (a)-(d).

The energy scale of the chemical potential $\mu$ and the pair potential $\Delta_0$ for the chiral $p$-wave SC and the chiral $p$-wave SC/QH insulator heterostructure is determined as follows.
For the chiral $p$-wave SC, we can consider arbitrary $\mu$ and $\Delta$ as long as the corresponding chirality is energetically favored (Fig.~\ref{fig:cihralp}) and $\Delta_0$ is sufficiently larger than that required for $\Delta_0$ in the heterostructure.
On the other hand, for the chiral $p$-wave SC/QH insulator heterostructure, we consider $\mu$ around the lowest and first Landau level ($\mu\sim\mu_\text{LLL},\mu_\text{1LL}$), and $\Delta_0$ smaller than the Landau level spacing $\mu_\text{1LL}-\mu_\text{LLL}$ (regions enclosed by dashed lines in Fig.~\ref{fig:cihralp}).

\subsection{Mixed-state chiral $p$-wave SC}

The BdG Chern number of the mixed-state chiral $p$-wave SC in \cite{PhysRevB.92.134519,PhysRevB.92.134520} has been evaluated far below $H_{c2}$ as we mentioned in Sec.~\ref{sec:chernnumber}.
Since strong magnetic field is necessary for the integer and fractional QH effects,
we examine whether the BdG Chern number of the mixed-state chiral $p$-wave SC is still even, even if the magnetic field is close to $H_{c2}$. 

\begin{figure}
    \centering
    \includegraphics[width=0.47\textwidth]{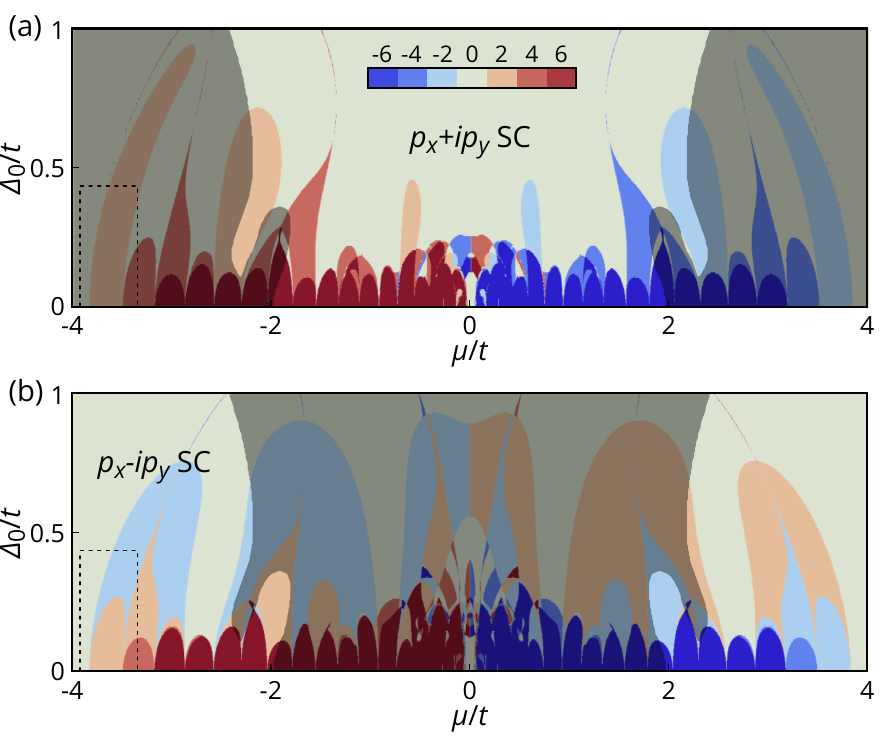}
    \caption{
        The Bogoliubov-de Gennes (BdG) Chern number of mixed-state (a) $p_x+ip_y$-wave and (b) $p_x-ip_y$-wave superconductors.
        Phases with BdG Chern number larger than 6 or less than -6 are filled by the same colors for simplicity.
        Unshaded regions show that the corresponding chirality is energetically favored.
        \label{fig:cihralp}
    }
\end{figure}

We numerically evaluate the BdG Chern number of mixed-state chiral $p$-wave SC close to $H_{c2}$ by (\ref{eq:tightbinding}) with (\ref{eq:pairpotential_tightbinding}) [Figs.~\ref{fig:cihralp} (a) and (b)].
The BdG Chern number is always even irrespective of the magnitude of $\Delta_0$ and $\mu$.
Notice that the BdG Chern number of the chiral $p$-wave SC without the vortex lattice is $\pm1$ when $\text{sgn}[C\mu]\lessgtr 0$, where $C$ is the chirality of the pair potential.
The energetically favored pairing chirality is shown by unshaded regions in Fig.~\ref{fig:cihralp}.
This indicates that the chirality can be tuned by the chemical potential.
Notice that the phase diagram around the $n+1$th Landau level of the $p_x+ip_y$-wave SC and that around the $n$th Landau level of the $p_x-ip_y$-wave SC have a similar structure within a small $\Delta$ albeit the difference of the BdG Chern number by 2.
This can be understood by an identity
\begin{align}
    \Delta_{nknk'}^{p_x-ip_y}=\Delta_{n+1kn+1k'}^{p_x+ip_y}
    \label{eq:nplusandn+1minus}
\end{align}
of the continuum model from (\ref{eq:pairpotential_cylinder_chiralps}).

\subsection{Mixed-state chiral $p$-wave SC/QH insulator}

Since we confirmed that the mixed-state chiral $p$-wave SC has even BdG Chern number $\mathcal{N}$, 
its heterostructure with a QH insulator can have odd $\mathcal{N}$ phases when a topological phase transitions occurs.
We will see that only the $p_x-ip_y$-wave pair potential can give rise to such a topological phase transition in the LLL.

We consider a QH insulator coupled with the proximity effect from a mixed-state chiral $p$-wave SC given by (\ref{eq:tightbinding}) with (\ref{eq:pairpotential_tightbinding}).
In particular, we focus on $\mu$ close to the LLL ($\mu_\text{LLL}$) and the first Landau level ($\mu_\text{1LL}$), and $\Delta$ smaller than $\mu_\text{1LL}-\mu_\text{LLL}$. 
\begin{figure}
    \centering
    \includegraphics[width=0.47\textwidth]{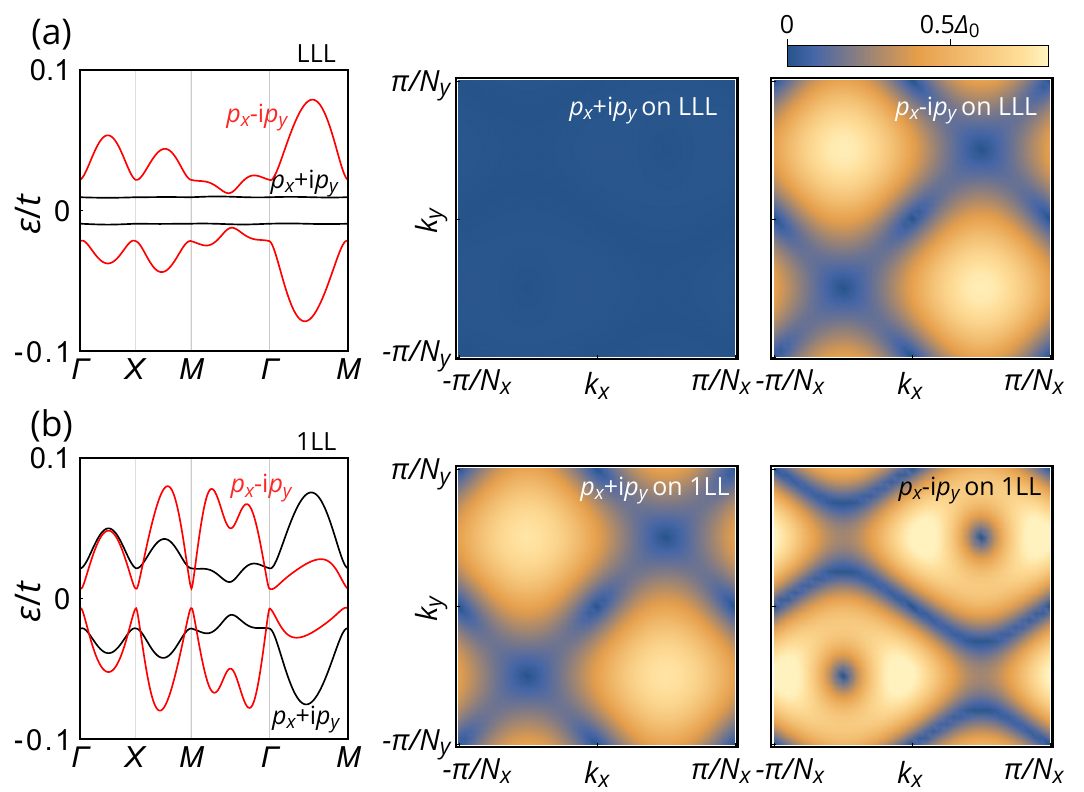}
    \caption{
        (a) Quasi-particle energy spectra with $p_x\pm ip_y$-wave pair potential (\ref{eq:pairpotential_tightbinding}) with $\Delta_0=0.1$ at $\mu_\text{LLL}$. The center and right figures are the expectation value of the pair potential with respect to the LLL states in the Brillouin zone.
        (b) The same plots as (a) at $\mu_\text{1LL}$.
        \label{fig:proximityband}
    }
\end{figure}
Let the momentum-space BdG Hamiltonian of (\ref{eq:tightbinding}) be denoted by $h(\bm{k})$.
The eigenenergies of $h(\bm{k})$ when $\mu=\mu_\text{LLL}$ are shown in Fig.~\ref{fig:proximityband} (a) left.
By diagonalizing the electron and hole parts of $h(\bm{k})$ separately, we obtain
\begin{align}
    &h(\bm{k})
    =
    \begin{pmatrix}
        h_0(\bm{k}) & \Delta(\bm{k})\\
        \Delta^\dagger(\bm{k}) & -h_0^\ast(-\bm{k})
    \end{pmatrix} \notag\\
    &\to
    \begin{pmatrix}
        \text{diag}[\epsilon_0(\bm{k})-\mu,\cdots] & \tilde{\Delta}(\bm{k})\\
        \tilde{\Delta}^\dagger(\bm{k}) & \text{diag}[-\epsilon_0(-\bm{k})+\mu,\cdots]
    \end{pmatrix},
    \label{eq:hetero_separatediagonalization}
\end{align}
where $\epsilon_0(\bm{k})$ is the LLL energy.
The (1,1)-component of $\tilde{\Delta}(\bm{k})$ in (\ref{eq:hetero_separatediagonalization}) is shown in Fig.~\ref{fig:proximityband} (a) center and right.
This quantity is the expectation value of the pair potential with respect to the LLL, and is the lattice counterpart of (\ref{eq:pairpotential_matrixelement_chiralp}) with $n=n'=0$.
The energy band remains flat by the $p_x+ip_y$-wave pair potential and the expectation value of the $p_x+ip_y$-wave pair potential is two orders of magnitude smaller than that of the $p_x-ip_y$-wave one.
These features cannot be seen in the first Landau level ($\mu=\mu_\text{1LL}$) [Fig.~\ref{fig:proximityband} (b)].
Notice that the coincidence of the $p_x-ip_y$-wave pair potential on the LLL and the $p_x+ip_y$-wave pair potential on the first Landau level [Fig.~\ref{fig:proximityband} (a) right and Fig.~\ref{fig:proximityband} (b) center] agrees with (\ref{eq:nplusandn+1minus}).
It is interesting to observe the quasi-particle energy spectra are also similar for $p_x-ip_y$ at the LLL and for $p_x+ip_y$ at 
the first Landau level [red curves in Fig.~\ref{fig:proximityband} (a) left and black curves in Fig.~\ref{fig:proximityband} (b) left].

\begin{figure}
    \centering
    \includegraphics[width=0.47\textwidth]{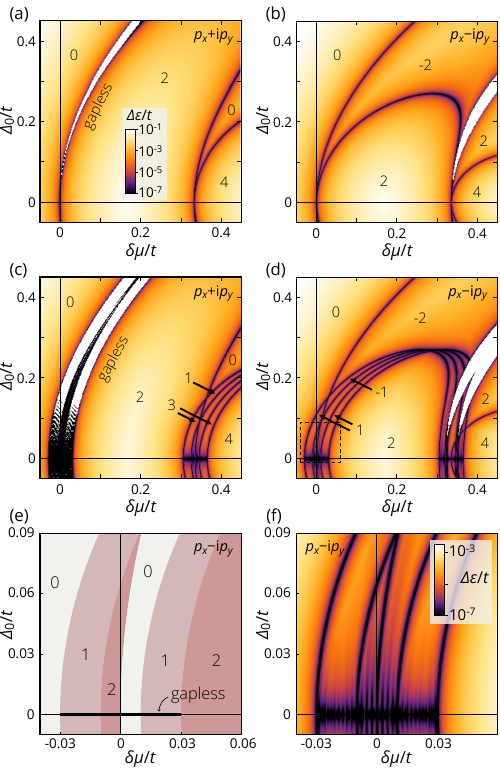}
    \caption{
        The phase diagrams of a quantum Hall insulator proximity-coupled with a $p_x+ip_y$- [(a), (c)] and $p_x-ip_y$-wave [(b), (d)] superconductor without [(a), (b)] and with [(c), (d)] a potential (\ref{eq:potential}) with $V_x=0.01$ and $V_y=0.02$.
        Color indicates the indirect energy gap $\Delta\epsilon$. $\delta\mu$ is the chemical potential measured from the lowest Landau level energy.
        The numbers inside the figures stands for the Bogoliubov-de Gennes Chern number.
        The indirect band gap is negative in white regions, that is, the lowest Bogoliubov quasi-particle band bents below the zero energy. 
        (e) and (f) are the phase diagram and the energy gap, respectively, of the region shown by a dashed line in (d). 
        \label{fig:phasediagram}
        }
\end{figure}

We numerically evaluate the BdG Chern number $\mathcal{N}$ of this system.
Here, we set $\mu$ to be around the lowest and first Landau level, and $\Delta$ to be up to the order of the energy gap between the two bands ($\sim 0.3t$) [which shares the same phase diagram as the chiral $p$-wave SC surrounded by dashed lines in Fig.~\ref{fig:cihralp} (a) and (b)].
Notice that the chemical potential $\mu$ of the QH insulator is irrelevant to that of the attached chiral $p$-wave SC in the previous subsection,
while the pair potential $\Delta$ induced in the QH insulator is assumed to be proportional to that of the chiral $p$-wave SC.
The energy gap and the BdG Chern number $\mathcal{N}$ are shown in Figs.~\ref{fig:phasediagram} (a) and (b) for $p_x\pm ip_y$-wave, respectively.
The appearance of the phase with $\mathcal{N}=-2$ is resulting from the proximity effect from the mixed-state $p_x-ip_y$-wave SC, 
while no new phases other than $\mathcal{N}=0$ and $2$ appear for the $p_x+ip_y$-wave case since the proximity effect does not work on the LLL. 

We can generate odd $\mathcal{N}$ phases between even $\mathcal{N}$ phases by adding external potentials \cite{PhysRevB.99.115427}. 
A topological phase transition accompanying more than $\mathcal{N}=1$ change can be split into those accompanying $\mathcal{N}=1$ change.
Specifically, for a topological phase transition from $\mathcal{N}=2$ to $-2$ in Fig.~\ref{fig:phasediagram} (b) around $\delta \mu/t\equiv(\mu-\mu_\text{LLL})/t\sim 0.2$, there are four momenta $\bm{k}=(0,0), (0,\pi/N_y), (\pi/N_x,0), (\pi/N_x,\pi/N_y)$ where the gap closes (Fig.~\ref{fig:proximityband}).
A potential
\begin{align}
    V(\bm{k}) = V_x \cos k_x N_x + V_y \cos k_y N_y
    \label{eq:potential}
\end{align}
with $V_x\neq V_y$ shifts the chemical potential at the four momenta differently.
As a result, topological phase transitions occur one by one by changing the chemical potential.

The phase diagrams with $V_x=0.01$ and $V_y=0.02$ are shown in Figs.~\ref{fig:phasediagram} (c) and (d) for $p_x\pm ip_y$-wave, respectively.
By the $p_x+ip_y$-wave pair potential, the LLL becomes dispersive but never be a topological SC. 
On the other hand, topological SC phases appear near the first Landau level.
By the $p_x-ip_y$-wave pair potential, topological SC phases appear even near the LLL.
These results agree with the prediction by the continuum model in the previous section.
The detailed phase diagram and the energy gap for the $p_x-ip_y$-wave case is shown in Fig.~\ref{fig:phasediagram} (e) and (f).
We can turn the LLL states into a topological SC by an infinitesimally small pair potential.

\section{Conclusion}
\label{sec:conclusion}

In this study, we consider SC/QH heterostructures and elucidated the role played by the vortex lattice in pairing bulk quantum Hall states.
Although the necessity of the vortex lattice was suggested even from the consideration of practical SC/QH heterostructures in Sec.~\ref{sec:generalhetero}, it turned out to be essential in inducing the bulk proximity effect irrespective of pairing symmetry.
By using the disk geometry, we showed that Cooper pairs with finite angular momenta generated by the vortex lattice can make pairs between time-reversal-symmetry broken states in the bulk QH state.
What we believe is essential in this conclusion is that we need to incorporate a magnetic field properly even in the SC side when we consider SC/QH heterostructures.
In this sense, it would be interesting to extend our result to a magnetic field away from $H_{c2}$, where the superconducting flux quantum is distributed sparsely and hence the pair potential can no longer be written by the LLL wavefunctions of Cooper pairs as in (\ref{eq:mixedstatepairpotential_disk}).

As an example, we considered a mixed-state chiral $p$-wave SC/QH insulator heterostructure, where both systems have distinct chiralities.
The presence of vortex lattice is not sufficient in pairing the LLL state; we need to align the relative chirality between the pairing and the quantum Hall states.
The pair potential by the opposite chirality SC is blocked in the LLL, while that by the same chirality gives rise to a topological phase transition to a topological SC with the aid of an external potential.
We confirmed these claims by both analytical calculations using a continuum model and numerical calculations using a tight-binding model.

Notice that we do not mean that our heterostructure is useful in constructing a topological superconductor since we made a topological SC from another topological SC.
We focus in particular on the physics underlying general SC/QH heterostructures rather than material science.
We believe that our study paves a way to engineering heterostructures with more exotic materials hosting non-Abelian anyons. 

Recently, the observation of the fractional QAH effect in moir\'{e} materials such as 
twisted MoTe$_2$ bilayers \cite{PhysRevLett.122.086402,Cai2023,Park2023,Zeng2023,PhysRevX.13.031037,Kang2024} 
and rhombohedral multilayer graphene \cite{Lu2024,PhysRevLett.133.206502,PhysRevLett.133.206503,PhysRevLett.133.206504,kwan2023moir} has attracted particular attention.
Heterostructures of these materials with SCs could realize SC/QH heterostructures without a magnetic field.
Though the electronic state of some of these materials are still under active discussion, it would be interesting to study the condition of the valid proximity effect in moir\'{e} QAH states.

\begin{acknowledgments}
    We thank Akira Furusaki, Masayuki Hashisaka, and Yukio Tanaka for discussions.
    The work is supported in part by
    JSPS KAKENHI Grant No.~JP24K06926, 
    No.~JP23K19036, 
    No.~JP25K17318, 
    No.~JP25H01250, 
    and No.~JP25H00613. 
\end{acknowledgments}

\appendix

\section{Matrix elements of mixed-state pair potentials}
\label{sec:pairpotential_disk}

Here we derive the matrix elements (\ref{eq:matrixelement_mixedstatespairpotential_disk}) and (\ref{eq:matrixelement_mixedstatechiralppairpotential_disk}) of the mixed-state $s$-wave and $p_x-ip_y$-wave pair potentials, respectively.
From (\ref{eq:pairpotential_matrixelement_sdisk}) and (\ref{eq:mixedstatepairpotential_disk}), one obtains
\begin{align}
    &\Delta_{nmn'm'}^s
    =
    \int d\bm{r}\phi_{nm}^\ast(\bm{r})\Delta(\bm{r})\phi_{n'm'}^\ast(\bm{r}) \notag\\
    &=
    \frac{(-1)^{n+n'}C_{m+m'}}{2\sqrt{\pi}\ell_B}
    \sqrt{\frac{n!n'!(m+m')!}{2^{m+m'}(n+m)!(n'+m')!}}I_{nmn'm'},
\end{align}
where
\begin{align}
    I_{nmn'm'}
    =
    \frac{2^{m+m'+1}}{(m+m')!}
    \int_0^\infty
    dx\,
    x^{m+m'}e^{-2x}L_n^m(x)L_{n'}^{m'}(x).
    \label{eq:pairpotential_matrixelement_coefficient}
\end{align}
Using $L_0^m(x)=1$ and $L_1^m(x)=m+1-x$, one can readily calculate (\ref{eq:pairpotential_matrixelement_coefficient}) for the first few cases as 
\begin{align}
    I_{0m0m'}&=1,\\
    I_{1m0m'}&=\frac{m-m'+1}{2}.
    \label{eq:i1m0m}
\end{align}
Notice that  $I_{0m1m'}$ can be obtained from (\ref{eq:i1m0m}) by an identity $I_{nmn'm'}=I_{n'm'nm}$.
These results are sufficient to evaluate (\ref{eq:matrixelement_mixedstatespairpotential_disk}) and (\ref{eq:matrixelement_mixedstatechiralppairpotential_disk}) 
using the $n=n'=0$ case of (\ref{eq:pairpotential_schiralprelation_disk}), that is,
\begin{align}
    \Delta_{0m0m'}^{p_x- ip_y}
    &=
    -\frac{i\hbar}{2\sqrt{2}\ell_B} 
    \left(\Delta_{1m-10m'}^s-\Delta_{0m1m'-1}^s\right).
\end{align}

For arbitrary $n$ and $n'$, one obtains
\begin{align}
    &I_{nmn'm'} 
    = \sum_{j=0}^n\sum_{k=0}^{n'}
    \binom{n+m}{n-j}
    \binom{n'+m'}{n'-k}
    \frac{(m+m'+1)_{j+k}}{(-2)^{j+k}j!k!},
\end{align}
where $(a)_n=\Gamma(a+n)/\Gamma(a)$ is the Pochhammer symbol,
by using an expression of the associated Laguerre polynomial 
\begin{align}
    L_n^m(x)
    =
    \sum_{j=0}^n(-1)^j
    \binom{n+m}{n-j}\frac{x^j}{j!}.
\end{align}

We anticipate that the absolute value of the coefficient $C_m$ is approximately independent of $m$ due to the following reason.
The integral of the squared amplitude of the pair potential over a sufficiently large region $S$ should be proportional to its area $|S|$ as the vortex lattice is uniformly distributed.
From (\ref{eq:vortexlattice_pairamplitude}),
\begin{align}
    \int_S d\bm{r} |\Delta(\bm{r})|^2
    =
    \sum_{m\ge 0}|C_m|^2
    \simeq
    \langle|\Delta(\bm{r})|^2\rangle|S|,
    \label{eq:integral_pairpotentialamplitude}
\end{align}
where $\langle\cdots\rangle$ denotes the spatial average.
Here, we consider a disk of radius $\sqrt{m_0}\ell_B$.
Since each term in (\ref{eq:mixedstatepairpotential_disk}) is distributed around a circle of radius $\sqrt{m}\ell_B$, the upper bound of the summation of (\ref{eq:integral_pairpotentialamplitude}) is approximately $m_0$ and its area is $|S|=\pi m_0\ell_B^2$.
Since this approximate relation holds for any (sufficiently large) $m_0$, one can conclude that $|C_m|\simeq \ell_B\sqrt{\pi\langle|\Delta(\bm{r})|^2\rangle}$, which does not depend on $m$.

\section{Mixed-state pair potential on a cylinder}
\label{app:cylinder}

We consider a continuum model of a mixed-state chiral $p$-wave SC/QH insulator heterostructure on a cylinder for comparison with a lattice model.
This argument will be proceeded basically in parallel with the previous studies with mixed-state $s$-wave superconductors \cite{PhysRevB.99.115427,PhysRevB.101.024516}.

The QH state wavefunction on a cylinder which is periodic in the $y$-direction with circumference $L_y$ is given under the Landau gauge $\bm{A}=(0,Bx)$, by \cite{Jain07}
\begin{align}
    &\phi_{nk}(\bm{r})
    =
    \frac{\left(a_k^\dagger\right)^n}{\sqrt{n!}}\phi_{0k}(\bm{r}) \notag\\
    &=
    \frac{e^{iky}}{\sqrt{L_y}}
    \frac{1}{\sqrt{2^nn!\ell_B\sqrt{\pi}}}
    e^{-(x-x_k)^2/2\ell_B^2}H_n\left(\frac{x-x_k}{\ell_B}\right),
    \label{eq:landaulevelwavefunction_cylinder}
\end{align}
where $n\in\mathbb{Z}_{\ge 0}$, $k=2\pi j/L_y\,(j\in\mathbb{Z})$, $a_k=[\ell_B\partial_x+(x-x_k)/\ell_B]/\sqrt{2}$, $x_k=-k\ell_B^2$, and $H_n$ is the Hermite polynomial.
The $s$-wave and chiral $p$-wave pairing Hamiltonians are given by (\ref{eq:pairinghamiltonian_s}) and (\ref{eq:pairinghamiltonian_chihralp}), respectively. 
The $s$-wave pairing Hamiltonian on the basis of (\ref{eq:landaulevelwavefunction_cylinder}) is given by
\begin{align}
    H^{s}_\Delta
    =
    \sum_{nkn'k'}\Delta^{s}_{nkn'k'}c_{nk\uparrow}^\dagger c_{n'k'\downarrow}^\dagger + \text{H.c.},
\end{align}
where
\begin{align}
    &\Delta^{s}_{nkn'k'}
    =
    \int d\bm{r}
    \phi^\ast_{nk}(\bm{r})
    \Delta(\bm{r})
    \phi^\ast_{n'k'}(\bm{r}).
\end{align}
Notice that if the $s$-wave pair potential is uniform [$\Delta(\bm{r})=\Delta_0$], 
the pair potential is nonzero only when $k'=-k$ and is given by
\begin{align}
    &\Delta^{s}_{nkn'-k}
    = 
    \Delta_0
    \sqrt{\frac{2^{n'}n!}{2^nn'!}}
    e^{-x_k^2/\ell_B^2}
    \left(\frac{x_k}{\ell_B}\right)^{n'-n}
    L_n^{n'-n}\left(\frac{2x_k^2}{\ell_B^2}\right),
\end{align} 
where $L_n^\alpha$ is the associated Laguerre polynomial.
Unlike the disk case, the number of induced pairs are not limited by a constant.
The strongest pairing is that between $k=0$ states, where $x_k=x_{-k}=0$ and are related by the time-reversal operation.
However, the matrix element decays exponentially as $x_k$ is far away from $x=0$. 
In this sense, the same conclusion as in the disk case holds, that is, a uniform $s$-wave pair potential does not induce an extensive number of Cooper pairs in QH states. 

The induced pair potential $\Delta(\bm{r})$ close to $H_{c2}$ is spanned by the LLL wavefunctions of Cooper pairs as \cite{tinkham2004introduction}
\begin{align}
    \Delta(\bm{r})
    =
    \sum_{j\in\mathbb{Z}}C_je^{iq_jy}e^{-(x-\tilde{x}_{q_j})^2/\ell_B^2},
    \label{eq:vortexlattice_pairamplitude}
\end{align}
where $\tilde{x}_q=-q\ell_B^2/2$. 
When the lattice vectors of the vortex lattice are $\bm{a}_1=(0,\Delta y)$ and $\bm{a}_2=(\Delta x,\Delta y\sin\theta)$, the momentum is given by $q_j=2\pi j/\Delta y\,(j\in\mathbb{Z})$.
The resulting matrix element of (\ref{eq:vortexlattice_pairamplitude}) is given by \cite{PhysRevB.99.115427,PhysRevB.101.024516}
\begin{align}
    \Delta^{s}_{nkn'k}
    =& 
    \sum_j \frac{(-1)^n C_j \delta_{q_j,k+k'}}{2^{n+n'}\sqrt{2n!n'!}}
    e^{-(x_k-x_{k'})^2/4\ell_B^2}
    H_{n+n'}\left(\frac{x_k-x_{k'}}{\sqrt{2}\ell_B}\right).
\end{align}

As in the disk geometry, the matrix elements of chiral $p$-wave pair potentials defined by 
\begin{align}
    H^{p_x\pm ip_y}_\Delta
    =
    \sum_{nkn'k'}\Delta^{p_x\pm ip_y}_{nkn'k'}c_{nk}^\dagger c_{n'k'}^\dagger + \text{H.c.}
\end{align}
are related to those of the $s$-wave one provided they share the same $\Delta(\bm{r})$.
Since the covariant derivative for electronic Landau level states of momentum $k$ is given by
\begin{align}
    e^{-iky}\left(-i\hbar D^e_\pm\right)e^{ik y}
    =
    \frac{\sqrt{2}i\hbar}{\ell_B}
    \left\{
    \begin{array}{ll}
        a_k^\dagger \,&(+)\\
     -a_k \,&(-)
    \end{array}
    \right.,
\end{align}
we obtain
\begin{align}
    &\Delta_{nkn'k'}^{p_x\pm ip_y}
    \equiv
    \frac{1}{2}
    \int d\bm{r}
    \psi_{nk}^\ast
    \frac{-i\hbar D_{\pm}^h\Delta+\Delta\left(-i\hbar D_{\pm}^e\right)}{2}
    \psi_{n'k'}^\ast \notag\\
    &=
    -\frac{i\hbar}{2\sqrt{2}\ell_B}
    \times\notag\\
    &
    \left\{
    \begin{array}{ll}
        -\sqrt{n}\Delta^s_{n-1kn'k'}+\sqrt{n'}\Delta^s_{nkn'-1k'}\,&(p_x+ip_y)\\[+3pt]
        \sqrt{n+1}\Delta^s_{n+1kn'k'}-\sqrt{n'+1}\Delta^s_{nkn'+1k'}\,&(p_x-ip_y)
    \end{array}
    \right..
    \label{eq:pairpotential_cylinder_chiralps}
\end{align}

Projecting onto the LLL ($n=n'=0$) and assuming (\ref{eq:vortexlattice_pairamplitude}), the pair potential is zero for the $p_x+ip_y$ wave while for the $p_x-ip_y$ wave,
\begin{align}
    \Delta^{p_x-ip_y}_{0k0k'}
    &=
    -\frac{i\hbar}{\sqrt{2}\ell_B}\Delta^s_{1k0k'}.
    \label{eq:pairpotential_lll}
\end{align}
Eq. (\ref{eq:pairpotential_lll}) is the same form as the one in a heterostructure of a mixed-state $s$-wave SC and a Rashba-coupled QH insulator \cite{PhysRevB.99.115427,PhysRevB.101.024516}.
A similar calculation on a sphere could also be done following \cite{PhysRevB.110.035147}.

\section{Pair potential in the tight-binding model}
\label{app:pplatice}

The coefficients $C_j$ in (\ref{eq:vortexlattice_pairamplitude}) are determined by imposing the periodicity of the vortex lattice.
Two lattice vectors of the vortex lattice are denoted by $\bm{a}_1=(0,\Delta y)$ and $\bm{a}_2=(\Delta x,\Delta y\sin\theta)$.
Notice that as two superconducting fluxes is equal to a flux quantum $h/e$, $\Delta x$ and $\Delta y$ are related by $\Delta x\Delta y=\pi \ell_B^2$ which is half of the area occupied by a flux quantum.
From $\bm{a}_1$, the momentum is given by $q_j=2\pi j/\Delta y\,(j\in\mathbb{Z})$.
The coefficient is given by $C_j=\Delta_0\exp[i\pi j^2\cos\theta]$ \cite{tinkham2004introduction}, that is,
\begin{align}
    \Delta(\bm{r})
    =
    \Delta_0
    \sum_{j\in\mathbb{Z}}e^{i\pi j^2\cos\theta}e^{2\pi ijy/\Delta y}e^{-\pi(x+j\Delta x)^2/\Delta x\Delta y}.
    \label{eq:vortexlattice_pairamplitude_beforerescaled}
\end{align}
Notice that the position of the vortices is $(\bm{a_1}+\bm{a_2})/2$ and their translations by $\bm{a}_1$ and $\bm{a}_2$.

A mixed-state pair potential on a square lattice model in Sec.~\ref{sec:lattice} is given by rescaling (\ref{eq:vortexlattice_pairamplitude_beforerescaled}).
Specifically, we consider an $N_x\times N_y$ magnetic unit cell in unit of the lattice constant of the tight-binding model, which contains two superconducting fluxes at the corners and the center of it [see Fig.~\ref{fig:pairpotential} (a)].
This gives $2\Delta x = N_x$, $\Delta y=N_y$, and $\cos\theta=1/2$.
Moreover, we translate the origin to $\bm{O}$ so that vortices in a magnetic unit cell are at  $(1/2,1/2)$ and $((N_x+1)/2,(N_y+1)/2)$.
This leads to an equation
\begin{align}
    \left(
        \frac{1}{2},\frac{1}{2}
    \right)
    =
    \bm{O}-\frac{\bm{a_1}+\bm{a_2}}{2}.
\end{align} 
From these result, we obtain (\ref{eq:pairpotential_tightbinding}).

\bibliography{chiralpiqh.bbl}

\end{document}